\documentclass{elsart}
\usepackage{epsfig}

\begin{document}

\begin{frontmatter}

\title{Correlations in the T Cell Response to Altered Peptide Ligands}

\author{Jeong-Man Park$^{1,2}$ and Michael W.\ Deem$^1$}

\address{
\hbox{}$^1$Department of Physics \& Astronomy and
Department of Bioengineering\\
Rice University, Houston, TX  77005--1892
}

\address{
\hbox{}$^2$Department of Physics\\
The Catholic University of Korea, Puchon 420--743, Korea
}

\begin{abstract}
The vertebrate immune system is a wonder of modern evolution.
Occasionally, however, correlations within the immune system 
lead to inappropriate  recruitment of preexisting T cells
against novel viral diseases.  We present a random energy theory
for the
correlations in the naive and memory T cell immune
responses.  
The non-linear susceptibility of the random energy
model to structural changes captures the correlations in the
immune response to mutated antigens.
We show how the sequence-level
diversity of the T cell repertoire drives the dynamics of
the immune response against mutated viral antigens.
\end{abstract}

\end{frontmatter}

\section{Introduction}

In this work, we develop a random energy model that allows us to address
limitations in the cellular immune system response
to mutable viral diseases.  
We focus on the T cell response,
with an emphasis on cytotoxic lymphocyte (CTL)
T cells, which bind the peptide-major
histocompatability class I (MHCI) complex
\cite{Perelson1997}.
These T cells eradicate other infected cells
through an interaction mediated by the binding of the
T cell receptor (TCR) to antigenic peptides derived from the
infectious disease.
The finite number of different T cells within an individual and
the existence of immune system memory makes the
immune system response to mutating viral diseases nontrivial.
We seek to understand how the 
T cell receptor repertoire changes within the immune system
in response to repeated exposure to evolving viral diseases.

The immune system has a mechanism for selecting 
among the many possible TCRs
those that best bind to the antigenic peptides.
TCRs are constructed from
modular elements, and each individual has an approximate diversity
of $2.5 \times 10^7$ different receptors \cite{Arstila}.
TCRs undergo rounds of selection for
increased affinity \cite{Blattman}.
TCRs do not undergo any further mutation during the
immune response.  Those TCRs that are stochastically
selected in the primary response to a disease become memory
cells \cite{Sourdive}.  While many details
affect the T cell selection process \cite{Lee,Lee2003}, selection for increased
affinity is thought to be a dominant factor
\cite{Kedl2003}.

Several limitations of the cellular immune system have been reported
that stem from the fact that a single T cell receptor
may bind multiple antigenic peptides.  This
cross-reactivity is studied in quantitative
molecular experiments through the use of
altered peptide ligands (APLs), peptides
that differ from the native ligand of the
TCR by one amino acid
 \cite{Allen}. 
We here develop a theory
of the T cell response to altered peptide ligands.
In section 2, we introduce the random energy 
model of the interaction between the TCR and the peptide-MHC complex.
In section 3, we describe the process by which T cells that bind
antigen are selected.  
In section 4, we show how experiments with altered peptide ligands
may be understood with the random energy model.
We discuss a critical point in the immune
response probability in section 5.  In section 6, we discuss correlations
that may arise in forward and reverse
experiments with altered peptide ligands.
We conclude in section 7.

\section{The Random Energy Model}
\begin{figure}[htbp]
\begin{center}
\epsfig{file=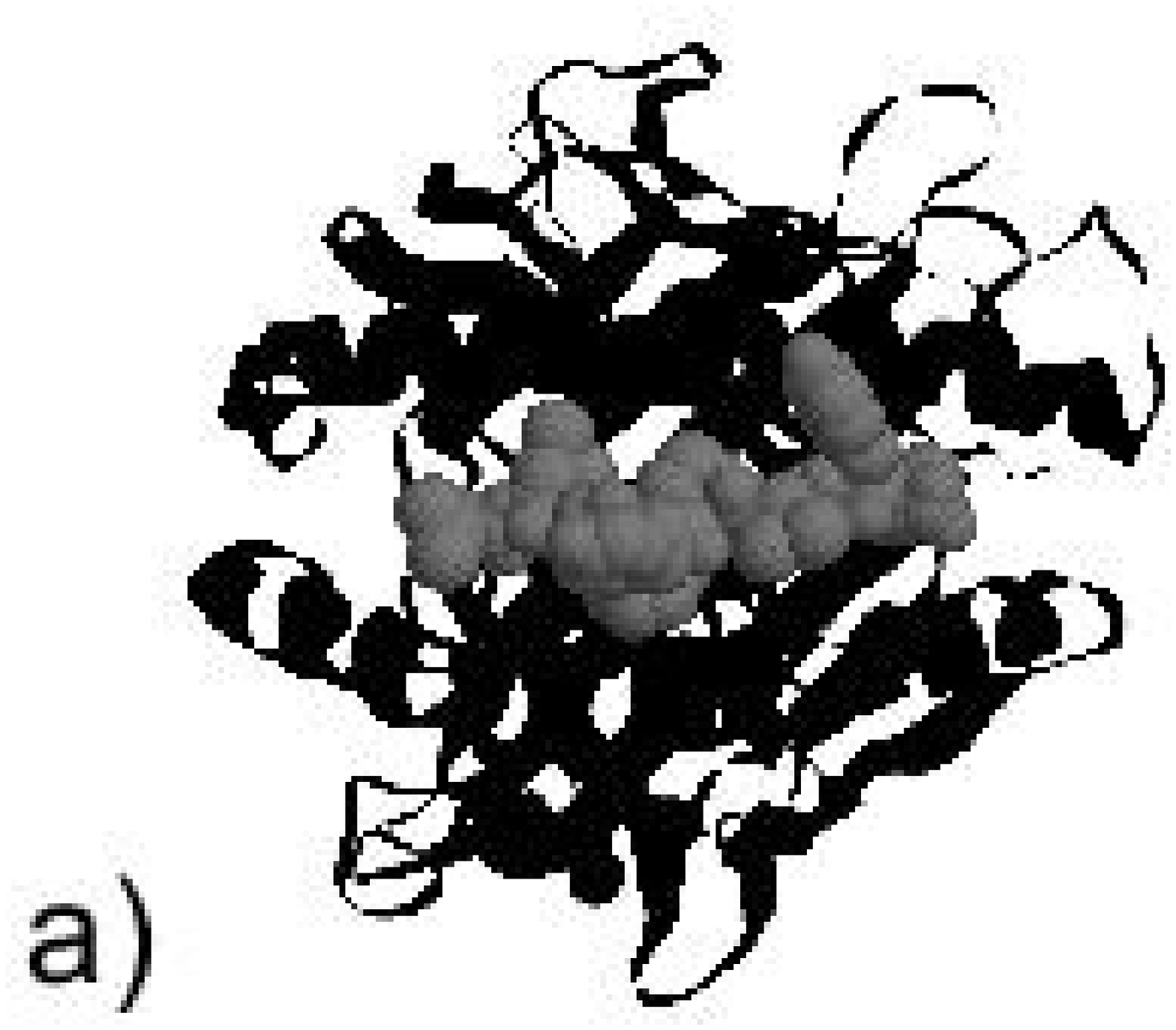,clip=,height=2in,angle=0} \\[0.5in]
\epsfig{file=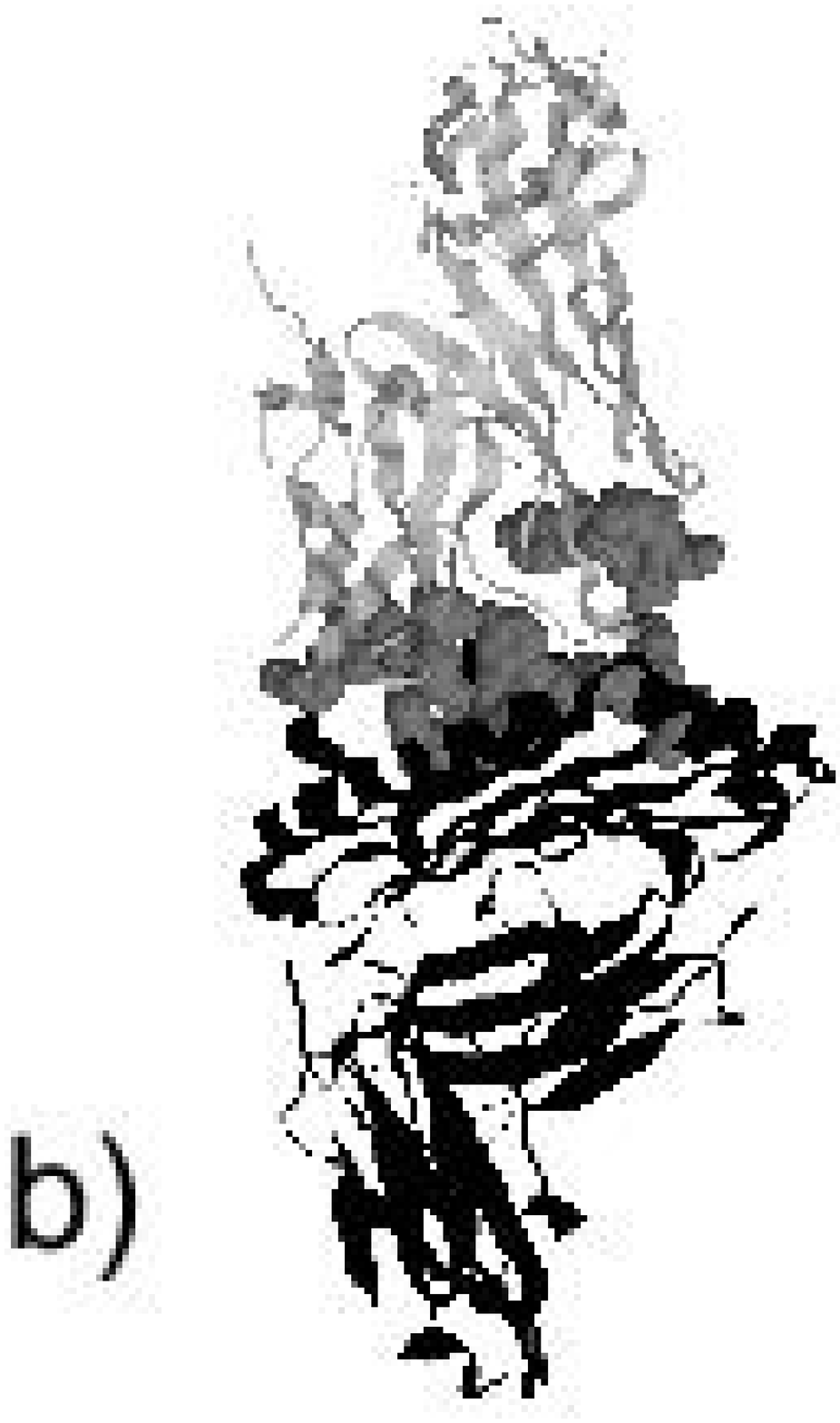,clip=,height=2in,angle=0}
\epsfig{file=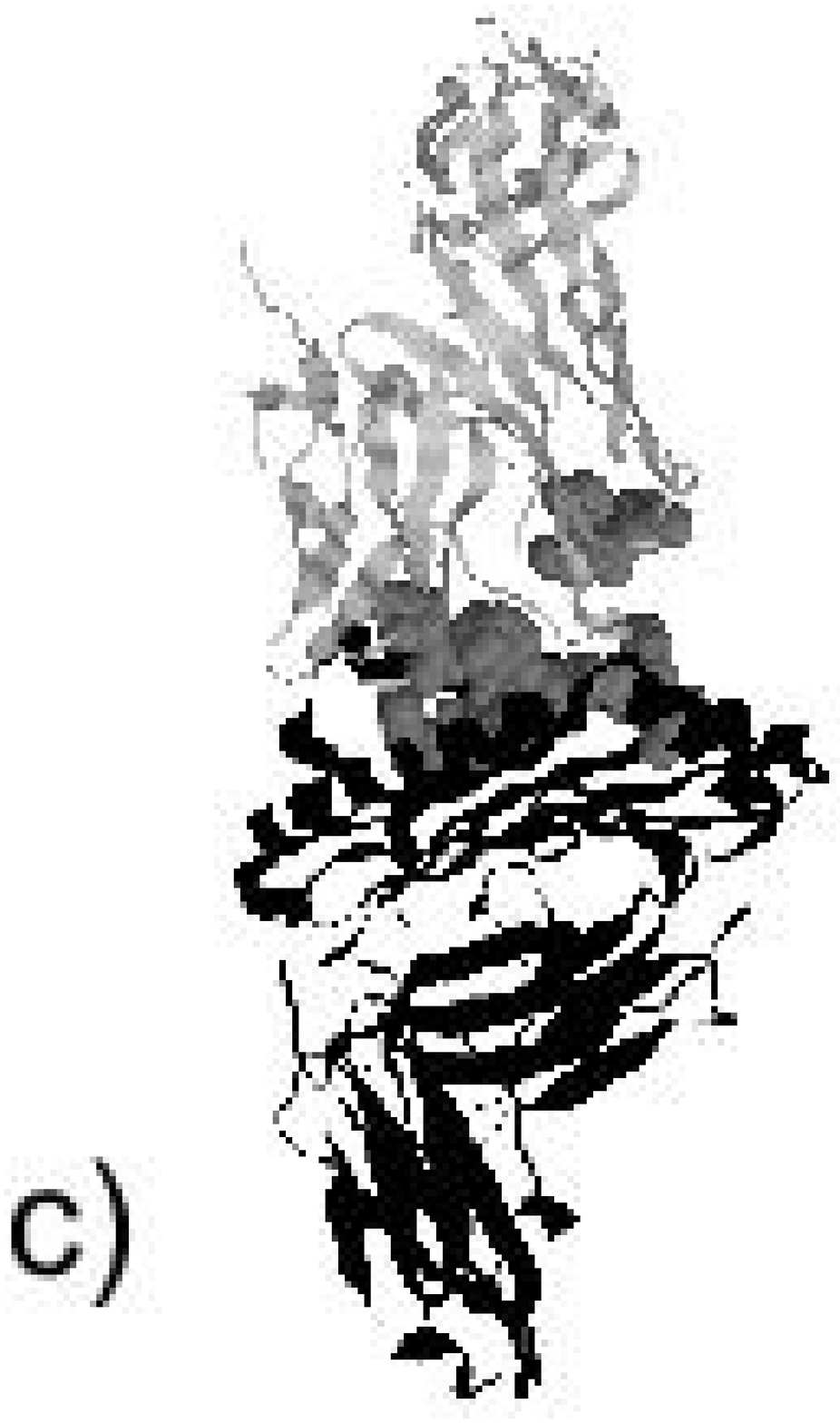,clip=,height=2in,angle=0}
\epsfig{file=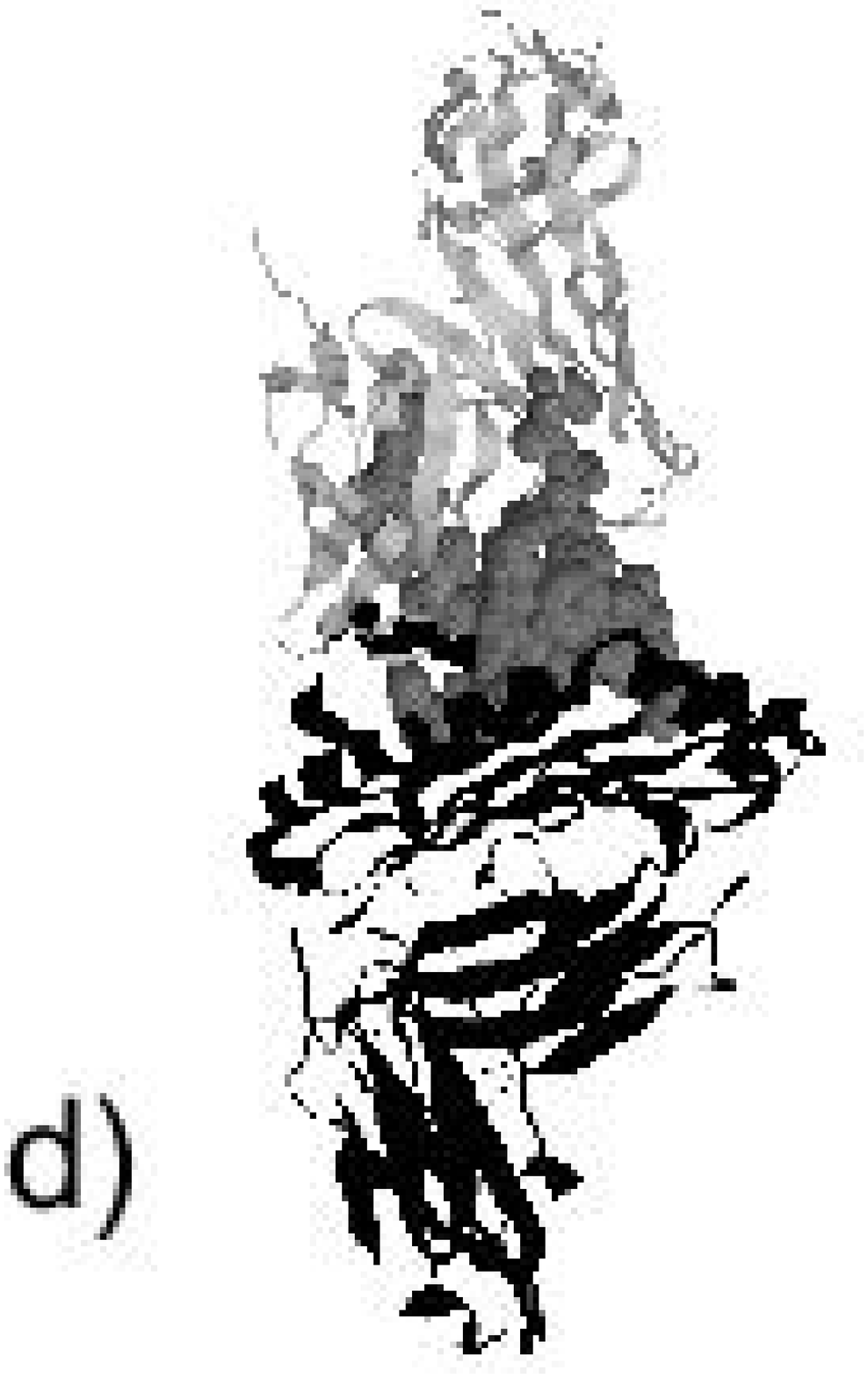,clip=,height=2in,angle=0}
\end{center}
\caption{
Figure of the interaction between the TCR and the
peptide-MHCI complex.  
The TCR consists of two domains, $\alpha$ and $\beta$, each of
which contains three complementary determining regions (CDRs) that
interact with the peptide-MHCI complex.  
The MHCI complex consists of
two domains, $\alpha$ and $\beta$, and the $\alpha$ domain
is comprised of two halves, $\alpha_1$ and $\alpha_2$.
a) Top view of the peptide-MHCI complex.
   The peptide (gray) sits like a ``hotdog in a bun'' in the
   $\alpha_1$ and $\alpha_2$ domains of the MHCI complex (black).
b) Side view of CDR 1 (gray) from TCR domain $\alpha$
   interacting with the first few amino acids of the peptide and the
   MHCI and of CDR 1 (gray) from TCR domain $\beta$ 
   interacting with the last few amino acids of the peptide and the MHCI.
   The TCR (light gray) sits on top of the peptide-MHCI complex.
   The peptide sites between the TCR and MHCI complex.
c) CDR 2 (gray) from TCR domain $\alpha$ 
   interacts with the first few amino acids of the peptide and the
   MHCI, and CDR 2 (gray) from TCR domain $\beta$ 
   interacts with the last  few amino acids of the peptide and the MHCI.
d) CDR 3 (gray) from both TCR domains
   interacts with the middle few amino acids of the peptide.
Atomistic details of this interaction are from an
X-ray crystal structure (\cite{Garboczi},
http://www.rcsb.org/pdb/, accession number 1AO7).
The interactions between the peptide and
the TCR are represented by the last two terms of Eq.\ 
\ref{eq:tcell}.  The TCR itself must fold, and these terms are
represented by the first two terms of Eq.\ \ref{eq:tcell}.
The interactions between the TCR and the MHCI complex have been integrated
out in the model.
}
\label{fig:interaction}
\end{figure}

The important molecular components of the T cell recognition event
are the peptide-MHCI complex and the T cell receptor (TCR),
as shown in Figure \ref{fig:interaction}.
Typically, the peptide ligand is on the order of 9 amino acids long.
The TCR variable region is composed of 54 amino acids, grouped into 6
subdomains.  The TCR sits atop the peptide-MHCI complex during recognition.
Recognition of
specific immunogenic peptides is due to the interaction between
the subdomains of the TCR and the peptide.  The majority
of the amino acids in the peptide are directly recognized by 1--4
amino acids of the TCR, with typically 3 amino acids in the
peptide being hot spots of the binding \cite{Garcia}.
The entire TCR variable region 
interacts relatively non-specifically with the
MHCI complex.
There is also a relatively generic interaction between the peptide
and the MHCI complex.

There are too many atoms within the TCR-peptide-MHCI complex
and too many TCRs within an individual to treat this problem in
full atomistic detail.
We therefore use a type of random energy
model to represent the interactions between
the TCR,  peptide, and MHCI complex.
This random energy builds a hierarchy of correlations into
Derrida's random energy theory
\cite{Derrida1980}.  The immune system being a real-time example
of an evolving system, the random energy approach is a
natural one \cite{Derrida1991}.
Our random energy model is a generalization of Kauffman's $NK$
model \cite{Kauffman} to include correlations due to protein
secondary structure \cite{Perelson,Bogarad}.  The model
introduced here to account for the T cell immune response contains
an additional term relative to
that which describes the B cell immune response
\cite{hayoun}.
In detail, our generalized $NK$ model for the T cell response
considers four different kinds of interactions:
interaction within a subdomain of the TCR $(U^{\rm sd})$,
interactions between subdomains of the TCR $(U^{\rm sd-sd})$,
indirect interactions between the TCR and the peptide $(U^{\rm pep-sd})$, and
direct binding interaction between the TCR and peptide $(U^{\rm c})$.
In this model, interactions between the peptide or
TCR and the MHCI have been integrated out.  As is typical in statistical
mechanics \cite{Chandler},
integrating out these terms produces random interactions
in the Hamiltonian, and we assume these to be roughly of the form of
the other intermolecular terms in our model.
 In the altered peptide ligand experiments
considered here, the peptides, whether original or altered,
are typically expressed by the same MHCI.
The first three terms of the present random energy model are
identical to those used to model
protein evolution \cite{Bogarad} and the B cell antibody response
\cite{hayoun}.  The parameters within the model have been
determined either by earlier work \cite{Kauffman,Perelson}
or by structural biology \cite{Bogarad}.
The energy function of the TCR with peptide-MHCI is
\begin{equation}
U=\sum_{i=1}^{M} U_{\alpha_{i}}^{\rm sd}
+\sum_{i>j=1}^{M} U_{ij}^{\rm sd-sd}
+ \sum_{i=1}^{M} U_i^{\rm pep-sd}
+\sum_{i=1}^{N_{\rm b}} \sum_{j=1}^{N_{\rm CON}} U_{ij}^{\rm c} \ ,
\label{eq:tcell}
\end{equation}
where $M=6$ is the number of TCR secondary structural subdomains,
$N_{\rm b}=3$ is the number of hot-spot amino acids that directly bind
to the TCR,
and $N_{\rm CON}=3$ is the number of 
T cell amino acids contributing directly to the binding of each 
peptide amino acid.  The subdomain energy $U^{\rm sd}$ is
\begin{equation}
U_{\alpha_{i}}^{\rm sd}= {1 \over \sqrt{M (N-K+1) } } \sum_{j=1}^{N-K+1}
\sigma_{\alpha_i} (a_j, a_{j+1}, \cdot \cdot \cdot, a_{j+K-1}) \ ,
\end{equation}
where $N=9$ is the number of amino acids in a subdomain,
and $K=4$ is the range of local interaction within a subdomain.
The prefactor for this term, and the other three terms in 
Eq.\ \ref{eq:tcell}, is chosen
so that random sequences produce a unit variance of this term.
Thus, each term in Eq.\ \ref{eq:tcell}
contributes roughly equal weight, a priori.
All subdomains belong to one of $L=5$ different types
(\emph{e.g.}, helices, strands, loops, turns, and others).
The quenched Gaussian random number $\sigma_{\alpha_i}$ is different
for each value of its argument for a given subdomain type, $\alpha_i$.
All $\sigma$ values in the model are Gaussian random numbers
and have zero mean and unit variance.  The sigma
values are different for each value of the argument, subscript,
or superscript.
The variable $\alpha_i$ defines the type of secondary structure for
the $i^{\rm th}$ subdomain, $1 \le \alpha_i \le L$.
The energy of interaction between secondary structures is
\begin{eqnarray}
U_{i j}^{\rm sd-sd}&=& \sqrt{2 \over D M (M-1)}
\nonumber \\
&& \times
\sum_{k=1}^{D}  \sigma^{k}_{i j} (a^{i}_{j_1},
\cdot \cdot \cdot, a^{i}_{j_{K/2}};
 a^{j}_{j_{K/2+1}},
\cdot \cdot \cdot, a^{j}_{j_K}) \ .
\nonumber \\
\end{eqnarray}
We set the number of interactions between secondary structures
at $D=2$, as the TCR-peptide-MHCI interaction
is a slightly less rough than is the antibody-antigen interaction
landscape where we have used $D=6$ \cite{hayoun}.
 Here $\sigma^{k}_{i j}$ and the interacting
amino acids, ${j_1, \cdot \cdot \cdot, j_K}$, are selected
at random for each interaction $(i, j, k)$.
The indirect interaction energy between the peptide and the
TCR is given by
\begin{eqnarray}
U_i^{\rm pep-sd} =
 \sqrt{1 \over D M}
\sum_{k=1}^{D} && \sigma^{k}_{i} (a^{\rm pep}_{j_1},
\cdot \cdot \cdot, a^{\rm pep}_{j_{K/2}};
 a^{i}_{j_{K/2+1}},
\cdot \cdot \cdot, a^{i}_{j_K}) \ .
\label{eq:tcell_p_sd}
\end{eqnarray}
Here $\sigma^{k}_{i}$ and the interacting
amino acids, ${j_1, \cdot \cdot \cdot, j_K}$, are selected
at random in the peptide and TCR subdomain
for each interaction $(i, k)$.
The chemical binding energy of each TCR amino acid to the peptide
is given by
\begin{equation}
U_{ij}^{\rm c}= \frac{1}{\sqrt{N_{\rm b} N_{\rm CON}}} \sigma_{ij}
              (a_{j_1}^{\rm pep}, a_{j_2})
\ .
\end{equation}
The contributing amino acids, $j_1, j_2$, and the
unit-normal weight of the binding, $\sigma_{ij}$, are chosen at random
for each interaction $(i,j)$, with $N_{\rm b}$ possible values for
$j_1$, and $N M$ possible values for $j_2$.

Although there are 20 different naturally occurring amino acids,
there are only roughly five distinct classes.  Mutations that
change an amino acid to another within the same class are
termed conservative, whereas mutations that change the class
of the amino acid are termed non-conservative.  This distinction is
significant because non-conservative mutations change the energy
landscape more dramatically than do conservative mutations.
Non-conservative mutations
lead to a factor of $2.34 \approx (1+1/2^2)^{1/2} / (1/2)$ 
greater energy change than do conservative
mutations, on average as estimated from statistically determined energy
values from the protein sequence alignment matrix
(PAM)
\cite{PAM}. To consider  all 20 amino acids within the random energy
model, Eq.\ \ref{eq:tcell}, and to consider
the differing effects of conservative and non-conservative mutations,
we set the random $\sigma$ for amino acid $i$
that belongs to group $j$ as $\sigma = w_j + w_i/2$, where the $w$ are
Gaussian  random numbers with zero average and unit standard deviation.
There are 5 groups, with 8 amino acids in the neutral and polar plus cystein
group, 2 amino acids in the negative and polar group,
3 amino acids in the positive and polar group, 4 amino acids in the
nonpolar without ring group, and 3 amino acids in the
nonpolar with ring group.  The results are not greatly sensitive to the
precise groupings of the amino acids.

The naive TCR repertoire is generated randomly from gene fragments.
This is accomplished by constructing the TCRs from subdomain pools.
Fragments for each of the $L$ subdomain types are
chosen randomly from 5 of the 100 lowest energy subdomain sequences.
This diversity mimics the known TCR diversity,
$(5 \times L)^M \approx 10^8$.  There have been recent suggestions that
the possible diversity of a given individual's TCR repertoire is
$\approx 10^{11}$, with the actual diversity expressed at any point in time 
being $\approx 10^8$ \cite{Kesmir2000}.  These numbers, which are very
much in line with the numbers for human antibodies,
would imply $L = 13$.  The results for the present study
are not greatly sensitive
to whether $L = 5$ or $L = 13$.

The binding constants between the peptide-MHCI ligand and the TCR
are calibrated by fixing the worst and geometric
mean of the binding constants evolved during a primary response
to be $3 \times 10^5$ l/mol and $3 \times 10^6$ l/mol,
respectively,
where the binding constant is related to the energy by
\begin{equation}
k = e^{a-b U} \ .
\label{eq:binding}
\end{equation}
That is, fixing the worst and geometric
mean of the binding constants to these values determines
the values of $a$ and $b$ for each instance of the random parameters
with in the generalized $NK$ model.
This approach leads
to the best binding constant being on average $3 \times 10^7$ l/mol.  These
binding constants are taken from experiment \cite{Schodin}
and are slightly smaller than those
for antibodies.  Whenever the peptide substrate is changed,
the constants $a$ and $b$ in Eq.\ \ref{eq:binding}
are reevaluated by comparison to the existing naive TCR repertoire.
All terms from Eq.\ \ref{eq:tcell}
are significant, because which of the TCRs from the
repertoire best bind depends on the identity of the peptide.

\section{The T Cell Maturation Process}

The T-cell-mediated response is driven by cycles of
concentration expansion and selection for better binding constants.
Selection processes are ubiquitous in nature \cite{Drossel2001},
although the exact mechanism of the T cell expansion during the primary
immune response remains elusive.
 It is clear
that the expansion of the T cells is non-linear \cite{Blattman}.
It also seems that, in most cases, there is competition among the
T cells for the presented antigen \cite{Kedl2003,Note1}.
Since the T cells do not mutate
during the primary response, our model of the primary response dynamics
to be detailed below
can be viewed as one particular 
description of the non-linear expansion of the naive
T cell repertoire.
The primary response increases the concentration of selected
TCRs by 1000 fold over 10 days, with a rough T cell doubling
time of one day.  The diversity of the memory
sequences is 0.5\% of that of the naive repertoire \cite{Arstila}.
The secondary response increases the concentration of the selected sequences
by 10 fold over a few days.  For humans, there are roughly $2.5 \times
10^7$ distinct T cell sequences 
at a copy number of $2.4 \times 10^4$ in the naive
repertoire and roughly $1.5 \times
10^5$ sequences at a copy number of $2 \times 10^6$ in the
memory repertoire \cite{Arstila}.  Typical best and worst binding
constants of the memory repertoire are $10^7$ l/mol and
$10^5$ l/mol, respectively \cite{Schodin}.

We implement a selection model of
the T cell immune system maturation.
Specifically, 10 rounds of selection are performed during the
primary response, with the top $x=58$\%
of the sequences chosen at each round. 
That is, the probability of picking one of the sequences
for the next round is
\begin{equation}
P_{\rm select} = 
\left\{
\begin{array}{ll}
\frac{1}{0.58 N_{\rm size}} ,& {\rm ~for~} U \le U_*\\
0,  &  {\rm ~for~} U > U_*
\end{array}
\right.
\end{equation}
where $U_*$ is the energy for which 58\% of the sequences lie below.
This equation is employed $N_{\rm size}$ times, where $N_{\rm size}$
is the diversity of the T cell repertoire, \emph{i.e.}\ the
number of distinct T cell present within the immune system,
to select randomly the $N_{\rm size}$ sequences for the next round.
This procedure mimics the
concentration expansion factor of $10^3 \approx 2^{10}$ in the primary
response and leads to 0.5\% diversity of the memory repertoire,
because $0.58^{10} \approx 0.5$\% and 10 days of doubling leads to
a concentration expansion of $2^{10} = 1024$.

As discussed in more detail in section 4, the quality of the
primary immune response is often measured experimentally by an \emph{in vitro}
or \emph{ex vivo} assay.
For an \emph{in vivo} secondary response, we
calculate the average binding constants
of the memory and naive responses, and whichever
is larger determines whether the response will be from the memory
or naive repertoires.  
In fact, the memory binding constant is multiplied
by a factor of 100, as memory T cells are present at higher concentrations
and more broadly present in the tissue than are naive T cells.
If the memory cells are used in 
the secondary \emph{in vivo} response, the top $x=58$\%
of the sequences are chosen, and 3 rounds of selection are performed
\cite{Blattman,Sourdive}.  This mimics the concentration factor of
$10 \approx 2^3$ during the secondary memory response.   
Conversely, if the naive cells are used in
the secondary
\emph{in vivo} response, the dynamics is identical to that of
the primary response.
For an \emph{in vitro} secondary response, in which memory T cells
are extracted and stimulated in an \emph{in vitro} experiment,
3 rounds of selection are performed, starting with exclusively
memory sequences.

Implementation of our theory proceeds by computational simulation of
the generalized $NK$ model.  First, the peptide and the altered peptide
ligands are created.  Then, the random terms such as the secondary
structural types of the subdomains, the $\sigma$ values, the
binding sites, and the interaction sites of the generalized
$NK$ model are determined.  Then the fraction and identity of the
T cell repertoire that responds well to the peptide ligand is identified.
Typically, 1 in $10^5$ T cells responds well to any particular ligand,
and one thus expects on the order of $10^8 / 10^5 = 1000$ T cells to
participate in the naive response.  Then the values of the constants
$a$ and $b$ in Eq.\ \ref{eq:binding} are calculated.
The primary response of 10 rounds of selection is then carried out.
Finally, the secondary response, either 10 rounds from the 
naive pool or 3 rounds from the memory pool, depending on the
relative binding constants for the altered peptide is carried out.

\section{Experiments with Altered Peptide Ligands}

A fundamental way to measure the correlation between immune
responses to related antigens is to use altered peptide ligands
(APLs).
In such experiments, an immune response is first generated to
the original peptide. 
The peptide antigen is then changed, typically either
conservatively or non-conservatively and only at one
amino acid position \cite{Zinkernagel}.  The immune response
to this altered peptide antigen is then measured.
Intuitively, one expects that if the peptide antigen is altered
only slightly, the immune response to the APL will be relatively
high, whereas if the peptide is altered significantly, the immune
response to the APL will be rather low.

Specific lysis is a measure of the probability that
an activated T cell will recognize an antigen presenting cell that
is expressing a particular peptide-MHCI complex.
This quantity is measured as a function of the effector to target ratio,
$E_0/T_0$, the ratio of the number of T cells to the number of
antigen presenting cells.  Each T cell typically expresses on the
order of $10^5$ identical TCRs, and each antigen presenting 
cell typically expresses on the order of $2 \times 10^4$
peptide-MHCI complexes \cite{Schodin}.
In typical experiments, the specific lysis is measured over a
4 hour time period \cite{Bachmann1997}.
Although T cells can each kill many targets \emph{in vivo}, 
the experimental specific lysis assay requires the number of
of T cells to be greater than the number of targets for detectable killing.
Lysis is a statistical event, being on average roughly proportional
to the probability of a T cell binding the target cell.
To calculate the specific lysis curve, we calculate the amount of
the target cells that are bound by all T cells. 
 We do this by considering
the equilibrium between the effector T cells and the target
antigen presenting cells:
\begin{equation}
E_i + T \rightleftharpoons E_i T, ~~~~~K_i \ .
\end{equation}
This equation implies $[ E_i T ] / ([E_i] [T]) = K_i$.
Introducing the 
amount of antigen presenting cells that are bound by T cell $i$, 
$\xi_i$, and noting that
the effector concentration is typically higher than the concentration
of antigen presenting cells,
we find
\begin{equation}
\xi_i = K_i E_i^0 (T_0 - \sum_i \xi_i) \ .
\end{equation}
Summing over all T cells, we find for the
total amount of antigen presenting cells
bound by any T cell:
\begin{equation}
L  = \frac{ \sum_i K_i E_i^0}{1 + \sum_i K_i E_i^0} \ .
\end{equation}
where $L = \sum_i \xi_i/T_0$ is
the specific lysis.
Writing this in the form
\begin{equation}
L = \frac{z E_0/T_0}{1 +z E_0/T_0} \ ,
\label{eq:L}
\end{equation}
we find $z = T_0 \sum_i K_i E_i^0 / E_0  = T_0 \langle K \rangle$.
Thus, the
competitive binding process for the antigen presenting cells
can be viewed as a Langmuir
adsorption isotherm of the T cells onto the
antigen presenting cells.

The binding constant $K_i$ is that between the two cells. Assuming that the
free energies of binding for each of the TCR/peptide-MHCI interactions
are approximately additive,
we find $K_i = 10^5 \times 2 \times 10^4 k_i$, where $k_i$ is the 
molecular binding constant between TCR $i$ and the peptide-MHCI complex.
Typical values of the target cell concentration are
$2 \times 10^3$ target cells in 200 $\mu$l of solution, or
$T_0 = 1.66 \times 10^{-17}$M \cite{Bachmann1997}.
We, thus, find 
\begin{equation}
z = \frac{\langle k \rangle}{3 \times 10^7} \ .
\label{eq:z2}
\end{equation}
Interestingly, the maximum value of the binding
constant in Eq.\ \ref{eq:binding} is of the same order
of magnitude as the denominator in Eq.\ \ref{eq:z2}.
For convenience, we have chosen them to be identical.
Adding in some cooperativity in the binding, \emph{i.e.}\ 
assuming that the free energies of binding for each of the TCR/peptide-MHCI
interactions are not entirely additive, 
will only change details such as the denominator
in Eq.\ \ref{eq:z2}.

At infinite dilution of the T cells, we find that the average
number of antigen presenting cells that are lysed  by one
T cell is $L T_0 / E_0 = z$.  The quantity $z$ is, therefore,
the average clearance probability of one T cell.
Since the T cell typically can lyse
no more than one target cell in specific lysis assays,
we modify the definition of the clearance probability as
\begin{equation}
z = \frac{1}{N_{\rm size}} \sum_{i=1}^{N_{\rm size}}
        \min(1, k_i/3 \times 10^7) \ .
\label{eq:z}
\end{equation}
This equation is implemented by introducing a lower cutoff
in the energy according to $3 \times 10^7 = \exp(a-b U_{\rm min})$.

The immune response probability is a measure of the probability
that the clearance probability is greater than 50\%.    In other
words, the immune response probability is the average of
$H(L - 1/2)$, where $H(x) = 0, x < 0$ and $H(x) = 1, x> 0$.
Both the specific lysis and the
immune response probability are averaged over many instances of the
random energy model, typically $10^4$.

Both specific lysis and immune response probability can be measured 
experimentally either
\emph{in vitro} or \emph{ex vivo}.  Typically these responses are measured
after the primary response to the original peptide ligand and
before a true secondary response to the altered peptide ligand.
Experimentally, the \emph{ex vivo} response is measured by
preparing mice, immunizing with the virus, and after 8--10 days
removing the spleen of the mouse.  The T cells from the spleen
are then challenged with antigen presenting cells expressing
the specific altered peptide ligand.  The spleen in this case
contains both memory and naive T cells.
For the \emph{ex vivo} APL response, we use $N_{\rm size}/2$ 
distinct memory cells and 
$N_{\rm size}$ distinct naive cells,
where $N_{\rm size}$ is the diversity of the naive T cell repertoire,
to calculate the observables
\cite{Arstila}.
Experimentally, the \emph{in vitro} response is measured by
removing the spleen of the mouse, immunizing the spleen cells with
the virus, and after 8-10 days challenging those spleen cells
with  antigen presenting cells expressing
the specific altered peptide ligand.   The spleen in this case
contains predominantly memory cells.
For the \emph{in vitro} APL response, we use exclusively
$N_{\rm size}$ memory cells to calculate the
observables.

\begin{figure}[htbp]
\begin{center}
\epsfig{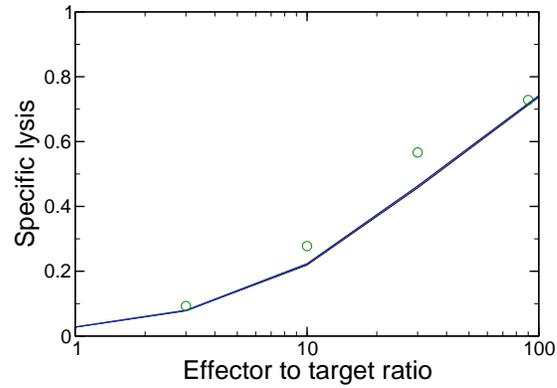} \\[0.5in]
\epsfig{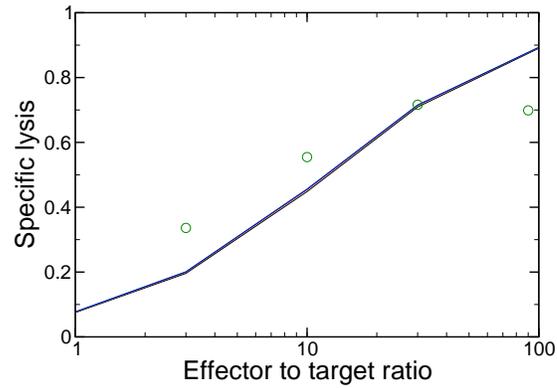}
\end{center}
\caption{
Specific lysis as a function of effector to target ratio for repertoire
sizes in the range $10^3$ to $10^6$ for non-conservatively altered
peptide ligands.
The theoretical curves overlap for all repertoire sizes.
  Results are for
\emph{ex vivo} (upper) and \emph{in vitro} (lower).
 Experimental data (circles) are from the LCMV
mouse models
of \cite{Bachmann1997} (\emph{ex vivo}) and \cite{Bachmann1998}
(\emph{in vitro}).
}
\label{fig:sl_non}
\end{figure}

We use this theory to analyze the correlations in the immune response
that are measured in altered peptide ligand experiments.
The non-linear susceptibility of the random energy
model to structural changes allows us to capture
the correlations in the
immune response to altered peptide antigens.
An extensive set of quantitative experiments on altered peptide ligands has
been carried out on the mouse model viral disease lymphocytic
choriomeningitis virus (LCMV). 
Due to the particularly strong immunogenicity of LCMV,
the memory T cell population is comprised essentially exclusively
of T cells from the primary LCMV response, in contrast to the
more typical  case where 1-10\% of the memory T cell population
is specific for a particular disease \cite{Barry}.
Shown in Figure \ref{fig:sl_non} is a
comparison between the measured and calculated \emph{ex vivo} and
\emph{in vitro} responses to altered peptide ligands with
a single non-conservative mutation.  Experiments have been carried out with
conservative mutations as well, and shown in Figure \ref{fig:sl_con} 
is a comparison between the measured and calculated \emph{ex vivo} and
\emph{in vitro} responses to altered peptide ligands with
a single conservative mutation.

The \emph{in vitro} response is always greater than the
\emph{ex vivo}, because a purely memory response is better than 
a naive response for these peptides altered by one amino
acid.  The responses are slightly superior for the conservative
mutations than for the non-conservative mutations,
because in the conservative case the altered peptide is more 
similar to the original peptide upon which the memory sequences were evolved.

Immune response probabilities have not been widely measured.
In a classic study, Klenerman and Zinkernagel found that
1 out of 7 non-conservatively altered peptides produced
a response to LCMV \cite{Zinkernagel}.  This data point is shown in 
Figure \ref{fig:irp_non}.  Also shown are the \emph{ex vivo} and
\emph{in vitro} immune response probabilities to altered peptide ligands with
a single non-conservative mutation.  Shown in Figure \ref{fig:irp_con}
are  the \emph{ex vivo} and \emph{in vitro} immune response probabilities
to altered peptide ligands with a single conservative mutation.

The \emph{in vitro} response remains superior to the
\emph{ex vivo} response, as the peptides with a
single amino acid mutation are best recognized by the
memory sequences.  In addition, the conservative
response is stronger than the non-conservative response,
because the conservatively mutated peptides are more similar
to the native target of the memory sequences than are the
non-conservatively mutated peptides.

\section{A Critical Point in the Immune Response Probability}
\begin{figure}[htbp]
\begin{center}
\epsfig{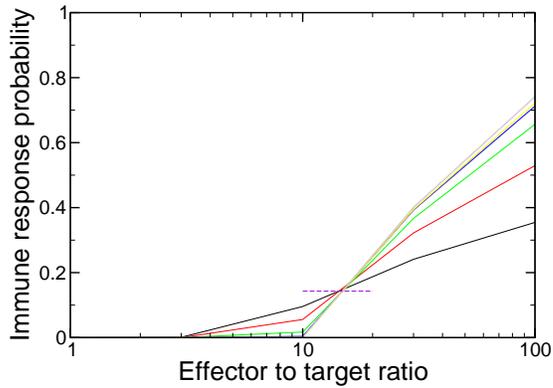} \\[0.5in]
\epsfig{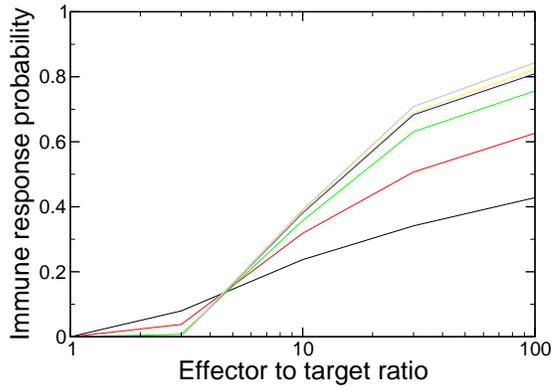}
\end{center}
\caption{
Immune response probability as
a function of effector to target ratio for repertoire
sizes of 
$10^3$, $3 \times 10^3$,
$10^4$, $3 \times 10^4$,
$10^5$, and $3 \times 10^5$
for non-conservatively altered peptide
ligands.  Larger repertoire sizes lead to a greater slope
at the inflection point of the curve.  Results are for
\emph{ex vivo} (upper) and \emph{in vitro} (lower).
Experimental datum (horizontal dashed line)
is taken from the LCMV mouse model of
\cite{Zinkernagel} (live mouse, \emph{in vivo}).
}
\label{fig:irp_non}
\end{figure}

\begin{figure}[htbp]
\begin{center}
\epsfig{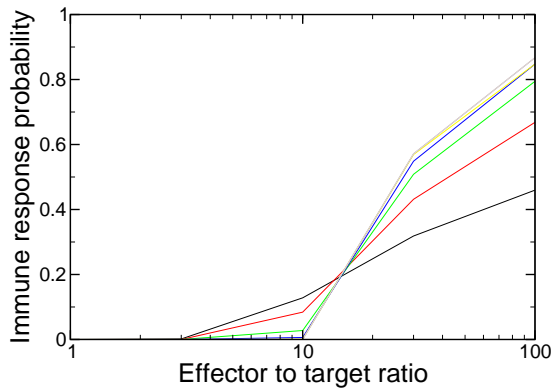} \\[0.5in]
\epsfig{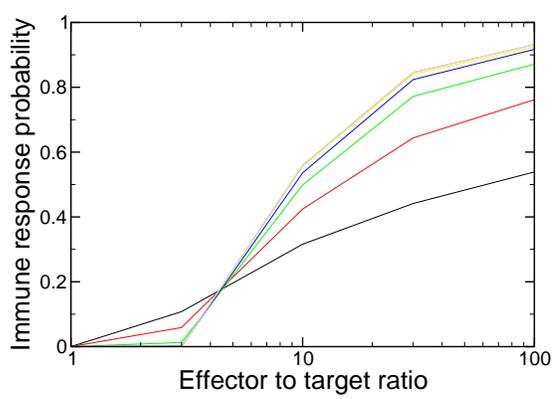}
\end{center}
\caption{
Immune response probability as a function of effector to target ratio
for conservatively altered peptide ligands.  
Repertoire sizes as in Figure \ref{fig:irp_non}.
Results are for \emph{ex vivo} (upper) and \emph{in vitro} (lower).
}
\label{fig:irp_con}
\end{figure}

It can be seen that larger repertoire sizes lead to a
sharpening of the immune response probability curve.
In fact, there is a critical point in our model which occurs
at $E_0/T_0 = 4.6$ for the \emph{in vitro} and
$E_0/T_0 = 16$ for the \emph{ex vivo} case.
The critical immune response probability lies in the
range 0.14--0.23.
The critical value of $E_0/T_0$ is shifted to a higher value for the
\emph{ex vivo} case because the average energies are not as favorable
in this case.  Randomness in the ensemble, due to the energy
fluctuations, causes the immune response probability in the
infinite repertoire limit to be a smooth curve, rather than a
step function.  Such randomness would correspond, for example,
to the variability in an individual's response to a variety of
disease strains or the variability in the response of a population of
individuals to a specific disease strain.  In our model, there is an
additional source of randomness for a finite repertoire size, the
inexact evaluation of the constants $a$ and $b$ in Eq.\ \ref{eq:binding}.
Assuming that the variation in $a$ is more significant and is Gaussian,
we can roughly say $z = \exp(a + \alpha \sigma - b U)$, where 
$\sigma$ is a Gaussian with unit variance and zero mean, 
$\alpha = O(1/\sqrt N_{\rm size})$
 is the rough size in the error of the calculation of
$a$, and $U$ is random with a certain probability distribution.
  We calculate the contribution of the randomness in $a$
to the immune response probability, $F$, as
\begin{eqnarray}
\frac{\partial F}{\partial \alpha} &=& \langle 
\frac{\partial H}{\partial L} \frac{\partial L}{\partial \alpha}
\rangle
\nonumber \\
&=& 
\langle
\delta(L - 1/2) 
\frac{\partial L}{\partial z}
\frac{\partial z}{\partial \alpha}
\rangle
\nonumber \\
&=& 
\frac{1}{4} \langle \sigma \vert L = 1/2 \rangle
\nonumber \\
&=& 
\frac{1}{4} \int d \sigma P(\sigma) \sigma \int du P(u) \delta(L - 1/2)
\nonumber \\
&=& 
\frac{1}{b} \int d \sigma P(\sigma) \sigma P(u^*)
\vert_{x \exp(a + \alpha \sigma - b u^*) = 1}
\end{eqnarray}
Assuming that the randomness in the energy is also Gaussian, we find
\begin{eqnarray}
F &=& F_0 - \frac{\alpha^2}{2 b^2 } x^{(-a + b \langle u \rangle)/(b^2 \chi)}
e^{-[(a-b \langle u \rangle)^2 + \ln^2 x]/(2 b^2 \chi)}
\nonumber \\ &&
\times
\frac{ a - b \langle u \rangle + \ln x }{\chi (2 \pi \chi)^{1/2}}
\end{eqnarray}
where $\chi = \langle (u - \langle u \rangle)^2 \rangle$.
The fixed point occurs for 
$x^* = \exp(-a + b \langle u \rangle)$
and $F^* = 1/2$.  The deviation of the immune response probability from
the infinite repertoire size limit is $O(\alpha^2) = O(1/N_{\rm size})$.
The shape of this curve is very similar to that seen in Figures
\ref{fig:irp_non} and \ref{fig:irp_con}.  If we make a more detailed
analysis that takes into account the negative asymmetry
of the probability distribution for $\ln z$,
which is due to the cutoff in
Eq.\ (\ref{eq:z}), we find that $F^*$ is lowered from 1/2.
The \emph{ex vivo} probability distribution for $\ln z$ is more
symmetric than is the \emph{in vitro} probability distribution, and so
the \emph{ex vivo} fixed point should be higher than the
\emph{in vitro} fixed point.  In addition, 
the conservative probability distribution for $\ln z$ has a smaller 
negative tail than does the non-conservative distribution, and so the
conservative fixed point should be higher than the non-conservative
fixed point.  These predictions are in agreement
with the observations in Figures
\ref{fig:irp_non} and \ref{fig:irp_con}.

\section{Correlation in Forward and Reverse APL Experiments}

\begin{figure}[htbp]
\begin{center}
\epsfig{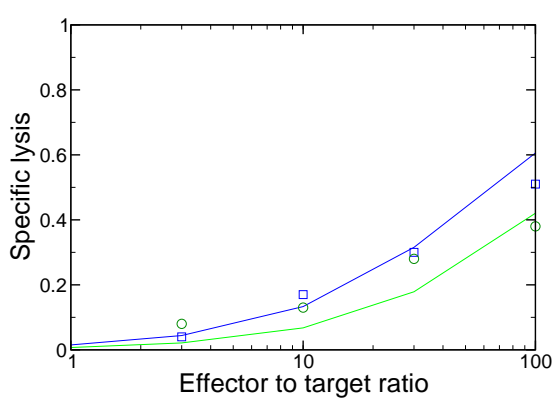}
\end{center}
\caption{
Representative specific lysis curves 
for a forward and backward altered peptide ligand experiment.
Data from experiments with
LCMV-WE original ligand and LCMV-8.7 altered peptide ligand (circles)
and LCMV-8.7 original ligand and LCMV-WE altered peptide ligand (squares)
from the \emph{in vitro} LCMV mouse model of \cite{Zinkernagel}.
}
\label{fig:asym}
\end{figure}

Using our random energy model, it is possible to examine how
correlations in the forward altered peptide ligand experiment
give rise to altered correlations in the reverse
altered peptide ligand experiment.
For example in Ref.\ \cite{Zinkernagel},
one set of experiments was performed with LCMV wild-type as the
original peptide and LCMV-8.7 as the altered peptide ligand,
and another set of experiments was performed with LCMV-8.7 as the
original peptide and LCMV wild type as the altered peptide ligand.
In Figure \ref{fig:asym}, we show these data
as well as a representative set of curves from our theory.

A more systematic way to study such forward and reverse experiments
is to look at the correlation between a response in the forward
original peptide ligand (OPL) $\to$ APL 
experiment and the reverse APL $\to$ OPL
experiment.  We introduce the correlation matrix
\begin{eqnarray}
C =
\left(
\begin{array}{cc}
\langle z_A z_B \rangle & \langle z_A (1-z_B) \rangle \\
\langle (1-z_A) z_B \rangle & \langle (1-z_A) (1-z_B) \rangle 
\end{array}
\right) \ .
\label{eq:matrix}
\end{eqnarray}
Here $z_A$ is the \emph{in vitro} clearance probability
of the APL in the forward experiment where the memory sequences were
evolved against the OPL, and
$z_B$ is the \emph{in vitro} clearance probability
of the OPL in the reverse experiment, where 
the memory sequences were evolved against the APL.
The averages are taken over instances of the random energy model.
Although the average response is the same for the forward and
reverse experiments, $\langle z_A  \rangle =
\langle z_B \rangle$, there may be a correlation between the two
responses.  To specify the notation, we note that the probability of
a forward response is given by $P(A) = \langle z_A  \rangle$ and that of
a reverse response by $P(B) = \langle z_B \rangle$.
The probability of both a forward and reverse response is
given by $P(A,B) = \langle z_A z_B \rangle  = C_{11}$.
The conditional 
probability that the reverse experiment is successful,
given that the forward experiment is successful, is
\begin{equation}
P(B \vert A) = \frac{P(A,B)}{ P(A)} = \frac{P(A,B)}{ P(B)} =
                \frac{C_{11}}{ \langle z_A \rangle}
 \ .
\end{equation}
From our model,
we find $\langle z_A \rangle = 0.113$ and $C_{11} = 0.01651$, and
thus $P(B \vert A) = 0.146$.  These values are not greatly sensitive to the
repertoire size.  The cross correlation is given by 
$( \Delta C)^2 = C_{11} - P(A) P(B) = 0.003797$.  The cross correlation
can alternatively be expressed as
\begin{equation}
\frac{P(B \vert A) }{ P(A)} = 1 + \left( \frac{\Delta C}{P(A)}\right)^2
\ .
\end{equation}
If there were no correlation between the two experiments, we would find
$P(B \vert A) / P(A) = 1$.  If there were complete correlation between
the experiments, $P(B \vert A) / P(A) = 1/P(A) = 8.87$.  In fact, we find
in our model that
 $P(B \vert A) / P(A) = 1.30$, showing a modest amount of correlation
between the OPL $\to$ APL and APL $\to$ OPL experiments.
Experimental measurements of this predicted
correlation would be very interesting.

\section{Summary}
The random energy model of the T cell immune system presented here allows
an investigation of the sequence-level evolution that occurs within the
T cell repertoire.  This is remarkable, given the complexity
underlying the immune response.
Correlations in the immune response to mutated antigens are
captured by the non-linear susceptibility of the underlying
random energy model to structural changes.
Specific lysis curves for the model virus
LCMV are well predicted by the theory.  The immune response probability is
shown to vary in a systematic fashion with the effector to target ratio,
and more experimental measurements of this quantity are needed.

The effective design of vaccines for viral diseases
requires some estimation of the likely escape mechanisms of the virus.  
For example, for rapidly mutating strains, a
highly-diverse, multicomponent vaccine may be necessary to halt progression
and transmission of the virus.  Conversely, for slowly evolving
strains, diversity within the vaccine may simply dilute the
conference of protective immunity.  
The approach taken here allows investigation and determination
of the fundamental qualitative and quantitative features that govern the
interaction between an effective multicomponent vaccine and the
variability of the virus.  The theory complements
and may provide some guidance to the long and difficult process of
experimental multicomponent vaccine development against escaping
viral diseases.


\section*{Acknowledgments}
It is a pleasure to acknowledge stimulating discussions with
Michael A.\ Barry.
This research was supported by the National Institutes of Health
and the National Science Foundation.

\bibliography{tcell}
\newpage

\clearpage

\begin{center}
{\bf Figures}
\end{center}       

\begin{figure}[htbp]
\begin{center}
\epsfig{file=fig3a.eps,clip=,height=2in} \\[0.5in]
\epsfig{file=fig3b.eps,clip=,height=2in}
\end{center}
\caption{
Specific lysis as a function of effector to target ratio for repertoire
sizes in the range $10^3$ to $10^6$ for conservatively altered peptide
ligands.  Results are for
\emph{ex vivo} (upper) and \emph{in vitro} (lower).
Experimental data (circles) are from the LCMV mouse models
of \cite{Zinkernagel} (\emph{ex vivo}) and
\cite{Martin} (\emph{in vitro}).
}
\label{fig:sl_con}
\end{figure}

\end{document}